\documentclass[preprint]{ptptex}
\notypesetlogo 
\newcommand{\cL}{\mbox{${\cal L}$}}
\newcommand{\cA}{\mbox{${\cal A}$}}
\newcommand{\cB}{\mbox{${\cal B}$}}
\newcommand{\cD}{\mbox{${\cal D}$}}
\newcommand{\cU}{\mbox{${\cal U}$}}
\newcommand{\ind}{\indent}
\newcommand{\nn}{\nonumber}
\newcommand{\be}{\begin{equation}}
\newcommand{\ee}{\end{equation}}
\newcommand{\beq}{\begin{eqnarray}}
\newcommand{\eeq}{\end{eqnarray}}
\newcommand{\ba}{\begin{array}}
\newcommand{\ea}{\end{array}}
 \def\bea*{\begin{eqnarray*}}
 \def\eea*{\end{eqnarray*}}
 \def\Vec#1{\mbox {\boldmath $#1$}}
 
 \newfont{\bg}{cmr10 scaled\magstep4}
\newcommand{\bigzerol}{\smash{\lower0.7ex\hbox{\bg 0}}}
\newcommand{\bigzerou}{%
 \smash{\lower0.5ex\hbox{\bg 0}}}
\newcommand{\bigzeroll}{\smash{\lower1.7ex\hbox{\bg 0}}}
\newcommand{\bigzerouu}{%
 \smash{\lower1.7ex\hbox{\bg 0}}}
\newcommand{\bigzerolll}{\smash{\lower2.5ex\hbox{\bg 0}}}
\newcommand{\bigzerouuu}{%
 \smash{\lower2.5ex\hbox{\bg 0}}}
\newcommand{\bigzeroL}{\smash{\lower0.01ex\hbox{\bg 0}}}
\title{%
A Dynamical Interpretation
of Connes' Unimodularity Condition
in Standard Model and Majorana Neutrino 
}
\vspace{5mm}
\author{%
  Katsusada {\sc Morita}$^{1}$ and
  Yoshitaka {\sc Okumura}$^{2}$
}
\inst{%
{\it  $^{1}$Department of Physics, 
  Nagoya University, Nagoya 464-8602, Japan,\\
 $^{2}$Department of Natural Sciences, 
  Chubu University, Kasugai 487-0027, Japan}
}
\vspace{5mm}
\abst{%
Standard model is minimally extended
using the unitary group 
$G'=U(3)\times SU(2)\times U(1)$
of Connes' color-flavor algebra.
In place of Connes' unimodularity
condition
an extra Higgs is assumed to
spontaneously break $G'$ down to
standard model
gauge group.
It is shown that the theory becomes anomaly-free
only if right-handed neutrino is present in each
generation.
It is also shown that
the extra Higgs gives rise to large Majorana mass
of right-handed neutrino
and
the model
predicts a new vectorial neutral current.
}
\begin{document}
\maketitle
\ind
The most promising idea of explaining
small neutrino mass is the sea-saw mechanism.\cite{1)}
The mechanism assumes the presence of right-handed
neutrino with normal Dirac mass and
large Majorana mass,
the mass eigenstates being Majorana neutrinos.
The purpose of the present note is 
to propose a way of introducing right-handed neutrino and the 
associated large energy scale in the standard model with a dynamical
interpretation of Connes' unimodularity condition.\cite{2)}
The latter
serves as a mathematical
restriction\cite{2),3)} to reduce the unitary group 
$G'=U(3)\times SU(2)\times U(1)$ of 
Connes' color-flavor algebra\cite{2)} to the standard model 
gauge group $G_{SM}=SU(3)\times SU(2)\times U(1)$. 
Our dynamical interpretation of the unimodularity condition
assumes extra Higgs singlet,
which triggers the symmetry breakdown, $G'\to G_{SM}$,
and generates an extra massive gauge boson singlet.
These singlets are expected not to affect the low energy phenomenology but 
the extra heavy gauge boson leads to a new vector-like neutral current.
It is incidentally shown that
right-handed neutrino is {\it necessary} to achieve
anomaly cancellation.
\\
\ind
Let us first summarize Connes' reformulation of the standard model
in a way convenient for later purpose.
The total fermion field is represented
as (bimodular) $4\times 4$ matrix-valued spinor
\begin{eqnarray}
\psi&=&\left(
    \begin{array}{cccc}
    l_L&q_L^r&q^b_L&q^g_L\\
    l_R&q_R^r&q^b_R&q^g_R\\
    \end{array}
    \right),
    \label{eqn:1}
\end{eqnarray}
where {\scriptsize$l_L=\left(
           \ba{l}
           \nu_e\\
           e\\
           \ea
           \right)_L,
           l_R=\left(
           \ba{l}
           \nu_{eR}\\
           e_R\\
           \ea
           \right),
           q_L=\left(
           \ba{l}
           u\\
           d\\
           \ea
           \right)_L,
           q_R=\left(
           \ba{l}
           u_R\\
           d_R\\
           \ea
           \right)$}
and $r,b,g$ are color indices.
For simplicity we consider only one generation. The 
gauge group in Connes' approach is taken as the unitary group of the 
color-flavor algebra\cite{2)}
\be
\cA_{\rm C}=C^\infty(M_4)\otimes({\Vec  H}\oplus{\Vec  C}\oplus M_3({\Vec  C})),\label{eqn:2}
\ee
where ${\Vec  H}$ denotes real quaternions, ${\Vec  C}$ complex field and
$M_3({\Vec  C})$ is the set of $3\times 3$ complex matrices so that
\begin{equation}
\cU(\cA_{\rm C})=\text{\rm Map}\,(M_4,U(3)\times SU(2)\times U(1))=G'.
\label{eqn:3}
\end{equation}
The gauge transformation is induced by the unitary restriction
of the algebra representation 
\begin{eqnarray}
\psi\to ^g\!\!\psi= g\psi G,
\label{eqn:4}
\end{eqnarray}
where 
\begin{eqnarray}
g&=&\left(
                               \begin{array}{cc}
                               g_L&0\\
                               0&g_R\\
                               \end{array}
                               \right),\;\;\;g_R=\left(
                               \begin{array}{cc}
                               u&0\\
                               0&u^{*}\\
                               \end{array}
                               \right),\;\;\;
G=\left(
    \begin{array}{cc}
    u^*&0\\
    0&V^T\\
    \end{array}
    \right).
\label{eqn:5}
\end{eqnarray}
Here $g_L\in SU(2)_L$, $u=e^{i\alpha}$, $\alpha$ being real, and 
$V\in \cU(M_3({\Vec  C}))$ are all local. On the other hand, the gauge 
transformation of Higgs doublet
  $h=\left(
  \begin{array}{cc}
  \phi_0^{*}&\phi_+\\
  -\phi_-&\phi_0\\
  \end{array}
 \right)$ 
reads \footnote{The hypercharge of Higgs $\phi$ is normalized to be $+1$,
leading to the choice $u=e^{i\alpha}$.}
\begin{eqnarray}
h&\to& ^gh=g_Lhg_R^{\dag}.
\label{eqn:6}
\end{eqnarray}
The spontaneous breakdown of symmetry,
$G_{SM}\to SU(3)\times U(1)_{em}$,
is given by
$g_L\to g_R$ since $\langle h\rangle=(v/\sqrt{2})1_2\ne 0$.
\\
\ind
Since det$\,g=1$ is 
obeyed, Connes' unimodularity condition\cite{2)} is formulated as
\begin{eqnarray}
\text{\rm det}\,G=1,
\label{eqn:7}
\end{eqnarray}
which reproduces correct hypercharge assignment of quarks (see
below).
Putting  $V=e^{i\beta}U$ with det$\;U=1$ and $\beta$ being real this implies
\begin{eqnarray}
-\alpha+3\beta=0.
\label{eqn:8}
\end{eqnarray}
However, the unimodularity condition (\ref{eqn:7}) is put 
by hand in this level. It is desirable to {\it derive} it 
by a dynamical mechanism. A dynamical derivation would not impose the
condition (\ref{eqn:8}) but rather regard $-\alpha+3\beta$
as an
{\it independent} gauge parameter.
To this end we 
first write Dirac Lagrangian invariant against the gauge transformation 
(\ref{eqn:4}):
\be
\cL_D=\text{\rm tr}\,
{\bar\psi}i\gamma^\mu[(\partial_\mu+\cA_\mu)\psi
+\psi\cB_\mu].
\label{eqn:9}
\ee
The gauge fields transform like
\begin{eqnarray}
\cA_\mu\to ^g\!\!\!\cA_\mu
&=&g\cA_\mu g^{\dag}+g\partial_\mu g^{\dag},\nn\\[2mm]
\cB_\mu\to ^g\!\!\cB_\mu&=&G^{\dag}\cB_\mu G-G^{\dag}\partial_\mu G.
\label{eqn:10}
\end{eqnarray}
Putting
\begin{eqnarray}
\cA_\mu&=&{\small\left(
                \ba{ccc}
                -ig_2/2\sum_{i=1}^3\tau_iA_\mu^i & &0\\
                0& &-ig_1/2{\small\left(
                                               \ba{cc}
                                               1&0\\
                                               0&-1\\
                                               \ea
                                               \right)}B_\mu\\
                \ea
                \right)},\nn\\[2mm]
\cB_\mu&=&{\small\left(
                \ba{cc}
                +ig_1/2B_\mu &0\\
                0& -ig_3/2\sum_{A=0}^8\lambda_A^TG_\mu^A\\
                \ea
                \right)},
\label{eqn:11}
\end{eqnarray}
where $\tau_i\;\;(i=1,2,3)$ are Pauli matrices, 
$\lambda_{A=a}\;\;(a=1,2,\cdots,8)$ Gell-Mann matrices with 
$\lambda_0=({g'}_3/g_3)1_3$,\footnote{From now on
we identify $G'=SU(3)\times SU(2)\times U(1)\times U(1)$.
The case ${g'}_3=g_3$ recovers the original
$G'$.}
the gauge transformation law (\ref{eqn:10}) for Abelian gauge fields
is cast into the form
\begin{eqnarray}
B_\mu&\to& ^g\!B_\mu=
B_\mu+(2/g_1)\partial_\mu\alpha,\nn\\[2mm]
G_\mu^0&\to&^gG_\mu^0=G_\mu^0+(2/g'_3)\partial_\mu\beta.
\label{eqn:12}
\end{eqnarray}
The unimodularity condition (\ref{eqn:7}) means the vanishing of the trace:
\begin{eqnarray}
\text{\rm tr}\cB_\mu=0\to
g_1B_\mu-3{g'}_3G_\mu^0=0.
\label{eqn:13}
\end{eqnarray}
Let us now derive (\ref{eqn:13}) as the low energy effective condition. This 
also means that the hypercharge gauge field is a mixture of $B_\mu$ and 
$G_\mu^0$. There are three neutral gauge bosons, $A_\mu^3,\;B_\mu$ and 
$G_\mu^0$ which can mix. Among them $A_\mu^3$ mixes at the electro-weak 
scale, while $B_\mu$ and $G_\mu^0$ are assumed to mix at much higher 
energy scale. Hence it is sufficient to consider the neutral coupling
involving $B_\mu$ and $G_\mu^0$ for our purpose:
\begin{eqnarray}
\cL_{NC;B,G^0}&=&
{\bar l}_Li\gamma^\mu(-ig_1/2){\small\left(
                                       \ba{cc}
                                       -1&0\\
                                       0&-1\\
                                       \ea
                                       \right)}
                                       B_\mu l_L
+{\bar l}_Ri\gamma^\mu(-ig_1/2){\small\left(
                                       \ba{cc}
                                       0&0\\
                                       0&-2\\
                                       \ea
                                       \right)}
                                       B_\mu l_R\nn\\[2mm]
&&+{\bar q}_Li\gamma^\mu(-ig'_3/2){\small\left(
                                       \ba{cc}
                                       1&0\\
                                       0&1\\
                                       \ea
                                       \right)}
                                       G_\mu^0q_L\nn\\[2mm]
&&+{\bar q}_Ri\gamma^\mu\left[(-ig_1/2){\small\left(
                                       \ba{cc}
                                       1&0\\
                                       0&-1\\
                                       \ea
                                       \right)}B_\mu
                                       +(-ig_3/2)\left(
                                       \ba{cc}
                                       1&0\\
                                       0&1\\
                                       \ea
                                       \right)G_\mu^0\right]
                                       q_R.
\label{eqn:14}
\end{eqnarray}
Writing ${g'}_3G_\mu^0=(1/3)g_1B_\mu+({g'}_3G_\mu^0-(1/3)g_1B_\mu)$
we separate the neutral coupling under consideration as
\begin{eqnarray}
\cL_{NC;B,G^0}&=&\cL_{NC;B}+\cL'_{NC;Z'},\nn\\[2mm]
\cL_{NC;B}&=&{\bar l}_Li\gamma^\mu(-ig_1/2)Y(l_L)B_\mu l_L
+{\bar l}_Ri\gamma^\mu(-ig_1/2)Y(l_R)B_\mu l_R\nn\\[2mm]
&&+{\bar q}_Li\gamma^\mu(-ig_1/2)Y(q_L)B_\mu q_L
+{\bar q}_Ri\gamma^\mu)(-ig_1/2)Y(q_R)B_\mu q_R,\nn\\[2mm]
\cL'_{NC;Z'}&=&{\bar q}i\gamma^\mu(-ig_0/2)\,Z'_\mu q,
\label{eqn:15}
\end{eqnarray}
where we have defined
\begin{eqnarray}
g_0\,Z'_\mu&=&{g'}_3G_\mu^0-(1/3)g_1B_\mu,
\label{eqn:16}
\end{eqnarray}
and $Y(f)$ is the hypercharge 2$\times$2 matrix
of fermions $f=l_L, l_R, q_L, q_R$. 
The unimodularity condition (\ref{eqn:13})
being imposed is tantamount to
putting $Z'_\mu=0$ by hand.
The resulting Lagrangian leads to
the correct hypercharge of fermions
with $B_\mu$ being the hypercharge gauge field
and $g_1$ being the hypercharge coupling constant.
This is what was achieved in Connes'
reformulation of standard model.\cite{2)}
\\
\ind
In contrast, we demand that both $Z'_\mu$ and its orthogonal
field $B'_\mu\propto (1/3)g_1G^0_\mu+g'_3B_\mu$ are dynamical
and the latter is to be identified with the hypercharge gauge field.
We write them as
\begin{eqnarray}
Z'_\mu&=&\cos{\delta}\,G_\mu^0-\sin{\delta}\,B_\mu,\nn\\[2mm]
B'_\mu&=&\sin{\delta}\,G_\mu^0+\cos{\delta}\,B_\mu,
\label{eqn:17}
\end{eqnarray}
where the mixing angle is given by
\begin{eqnarray}
\tan{\delta}=\frac {g_1}{3{g'}_3}.
\label{eqn:18}
\end{eqnarray}
They transform like
\begin{eqnarray}
Z'_\mu&\to& ^g\!Z'_\mu=
Z'_\mu+(2/g_0)\partial_\mu(\beta-\alpha/3)
=Z'_\mu-(2/g_1\sin{\delta})\partial_\mu(\alpha-\gamma),\nn\\[2mm]
B'_\mu&\to& ^g\!B'_\mu=B'_\mu+(2/g'_1)\partial_\mu\gamma,\hspace{1cm}
\gamma=\alpha+3\sin^2{\delta}(\beta-\alpha/3)
\equiv \alpha+\gamma'.
\label{eqn:19}
\end{eqnarray}
To make $Z'_\mu$ heavy but leave $B'_\mu$ massless
we assume
the presence of
an extra Higgs transforming like
\begin{eqnarray}
\Phi\to
^g\!\!\Phi=e^{2i(\alpha-\gamma)}\Phi.
\label{eqn:20}
\end{eqnarray}
Nonvanishing vacuum expectation value
$\langle\Phi\rangle\ne 0$
leaves $\alpha=\gamma$ unbroken.
The Lagrangian of the extra Higgs singlet is given by
\begin{eqnarray}
\cL_{\Phi}&=&(D_\mu\Phi)^{\dag}D^\mu\Phi
-\frac {\lambda'}4({\Phi}^{\dag}\Phi-\frac{{v'}^2}2)^2,\;\;\;
D_\mu\Phi=(\partial_\mu+ig_1\sin{\delta}\,Z'_\mu)\Phi,
\label{eqn:21}
\end{eqnarray}
which gives rise to the extra gauge boson 
mass,
$M_{Z'}^2=g_1^2\sin^2{\delta}{v'}^2,$
leaving $B'_\mu$ massless as desired.
By inserting 
the inverse of (\ref{eqn:17}) 
\begin{eqnarray}
G_\mu^0&=&\cos{\delta}\,Z'_\mu+\sin{\delta}\,B'_\mu,\nn\\[2mm]
B_\mu&=&-\sin{\delta}\,Z'_\mu+\cos{\delta}\,B'_\mu,
\label{eqn:22}
\end{eqnarray}
into the piece $\cL_{NC;B}$ and defining 
\begin{eqnarray}
{g'}_1=g_1\cos{\delta},
\label{eqn:23}
\end{eqnarray}
the piece $\cL_{NC;B}$ contains a part, $\cL_{NC}$, of the standard model 
Lagrangian with the hypercharge gauge field $B'_\mu$
and $g'_1$ being the 
hypercharge coupling constant. The remaining part 
can be combined with $\cL_{NC;Z'}$ to yield an additional neutral coupling, 
$\Delta'\cL_{NC}$, which involves only the gauge field $Z'_\mu$. 
The low energy standard model structure emerges if we let $Z'_\mu$ 
and $\nu_R$ be heavy enough to be unobservable at low energy. 
We shall see below that
$\Phi$ generates Majorana mass of order $v'$ for $\nu_R$,
and, hence,
we assume $v'\gg v$ so that the gauge 
boson $Z'$ effectively decouples from 
and $\nu_R$ does not appear in
the low energy spectrum. In this sense 
the unimodularity condition (\ref{eqn:13}) is regarded as the low energy 
condition.
Using the 
standard Higgs mechanism we obtain ($A^3_\mu$ is now included)
\begin{eqnarray}
\cL_{NC}&=&eA_\mu J_{em}^\mu+\frac {g_2}{\cos{\theta_W}}
Z_\mu J_Z^\mu,\nn\\[2mm]
\Delta'\cL_{NC}&=&
-\frac{g_1\sin{\delta}}2\,Z'_\mu J'_{Z'}{}^\mu-\frac {ig_0}2{\bar q}i\gamma^\mu\,Z'_\mu q,
\label{eqn:24}
\end{eqnarray}
where $\cL_{NC}$ has the conventional form with
$B'_\mu$ being the hypercharge gauge field and
$g'_1$ the hypercharge gauge coupling constant
and
\begin{eqnarray}
J'_{Z'}{}^\mu&=&{\bar l}_L\gamma^\mu Y(l_L)l_L
+{\bar l}_R\gamma^\mu Y(l_R)l_R+
{\bar q}_L\gamma^\mu Y(q_L)q_L
+{\bar q}_R\gamma^\mu Y(q_R)q_R.
\label{eqn:25}
\end{eqnarray}
Note that we have invoked the standard Higgs mechanism by employing the 
Lagrangian
\begin{eqnarray}
\cL_H&=&\text{\rm tr}\left[
(\cD_\mu h)^{\dag}\cD^\mu h-\frac \lambda 4(h^{\dag}h-\frac{v^2}21_2)^2\right],
\label{eqn:26}
\end{eqnarray}
where Higgs field $h$ is assumed to transform under gauge transformation
\begin{eqnarray}
h\to ^g\!\!h=g_Lhg_R^{\dag}\left(
                                \ba{cc}
                                e^{-i\gamma'}&0\\
                                0&e^{i\gamma'}\\
                                \ea
                                \right)\equiv g_Lhg'_R{}^{\dag},\;\;\;
g'_R=\left(
   \ba{cc}
   e^{i\gamma}&0\\
   0&e^{-i\gamma}\\
   \ea
   \right),
\label{eqn:27}
\end{eqnarray}
with $\gamma'=\sin^2{\delta}(3\beta-\alpha)$ and
$\gamma=\alpha+\gamma'$. The covariant derivative $\cD_\mu h$
is determined from (\ref{eqn:27}) with
the gauge transformation property
(\ref{eqn:19}).
The replacement of $g_R$ with $g'_R$ comes from the requirement
that $B'_\mu$ but not $B_\mu$
be the hypercharge gauge field
and $g'_1$ but not $g_1$ be the hypercharge
coupling constant.
The spontaneous symmetry breakdown, $G'\to SU(3)\times U(1)_{em}$,
is given by $g_L\to g'_R$ and $\gamma\to\alpha$.
\\
\ind
The above gauge transformation
of standard Higgs modifies that of fermions
in order to preserve gauge invariance of Yukawa coupling.
The modified gauge transformation is
\beq
l_L&\to& ^gl_L=e^{-i\alpha}g_Ll_L,\;
\nu_R\to ^g\nu_R=e^{-i\alpha+i\gamma}\nu_R,
e_R\to ^g\!e_R=e^{-i\alpha-i\gamma}e_R,\nn\\[2mm]
\phi&\to& ^g\!\phi=e^{i\gamma}g_L\phi,\nn\\[2mm]
q_L&\to& ^gq_L=e^{i\beta}g_LUq_L,
u_R\to ^gu_R=e^{i\beta+i\gamma}Uu_R,\;
d_R\to ^gd_R=e^{i\beta-i\gamma}Uu_R.
\label{eqn:28}
\eeq
This is equivalent to replacing $g_R$ in (\ref{eqn:4}) and (\ref{eqn:5})
with $g'_R$.
It is apparent that triangle anomaly
arising from lepton doublet running in the loop
does not cancel that of quark doublet unless
we perform the
following {\it additional} gauge transformation
\begin{eqnarray}
q\to ^g\!\!q=e^{-i(\beta-\alpha/3)}q,
\label{eqn:29}
\end{eqnarray}
so that the quark sector
in (\ref{eqn:28}) is modified as
\beq
&&\!\!\!
q_L\to ^g\!q_L=e^{i\alpha/3}g_LUq_L,
u_R\to ^g\!u_R=e^{i\alpha/3+i\gamma}Uu_R,
d_R\to ^g\!d_R=e^{i\alpha/3-i\gamma}Uu_R.
\label{eqn:30}
\eeq
The gauge transformation (\ref{eqn:29}) introduces
additional gauge interaction
which is just negative of $\cL'_{NC;Z'}$ in (\ref{eqn:15}). 
\\
\ind
The gauge transformation of leptons given by (\ref{eqn:28})
and that of quarks given by (\ref{eqn:30})
finally determine the gauge interaction of fermions.
By investigating non-safe triangle diagrams,
$B^3, SU(3)B'{}^2,$\\
$SU(3)SU(2)B', SU(2)B'{}^2, [SU(3)]^2B', [SU(3)]^2B,$
it can be shown that the theory is anomaly-free.
In particular,
$\nu_R$ participates in anomaly cancellation
in a non-trivial way.
\\
\ind
The neutral coupling involving $B$ and $B'$
is now given by
\begin{eqnarray}
\cL_{NC;B,G^0}+\Delta\cL_{NC;Z'}
&=&\cL_{NC;B'}+\cL_{NC;Z'},\nn\\[2mm]
\cL_{NC;Z'}&=&(g_1\sin{\delta}/2)Z'_\mu J_{Z'}^\mu,\;\;\;
J_{Z'}^\mu=[{\bar l}\gamma^\mu\,l
-(1/3){\bar q}i\gamma^\mu\,q],
\label{eqn:31}
\end{eqnarray}
where $\cL_{NC;B'}$ is the same as $\cL_{NC;B}$
given by (\ref{eqn:15}) with $B\to B'$ and $g_1\to g'_1$
and the extra coupling $\Delta\cL_{NC;Z'}$,
which comes from non-vanishing $\gamma'=
\gamma-\alpha=3\sin^2{\delta}(\beta-\alpha/3)$
associated with the gauge field $Z'_\mu$,
see (\ref{eqn:19}), is given by
\beq
\Delta\cL_{NC;Z'}=(ig_1\sin{\delta}/2){\bar l}_Ri\gamma^\mu Z'_\mu
{\small\left(
\ba{cc}
-1&0\\
0&1\\
\ea
\right)}l_R
+(l_R\to q_R).
\label{eqn:32}
\end{eqnarray}
\ind
Comparing (\ref{eqn:20}) and $\nu_R$ transformation property
of (\ref{eqn:28}) we also have the invariant coupling
\begin{eqnarray}
\nu_R^TC\Phi\nu_R+{\bar\nu}_R^TC\Phi^{\dag}{\bar\nu}_R,
\label{eqn:33}
\end{eqnarray}
which generates Majorana mass of order $v'$ for $\nu_R$.
Such Yukawa couplings are absent for other fermions.
Since neutrino possesses Yukawa coupling generating
normal Dirac mass,
the sea-saw mechanism works in this model without invoking
GUT.
\\
\ind
Unfortunately, the new energy scale $v'$ is quite arbitrary except that it 
must be very large compared with $v$ so as not to conflict with the present 
experiment. Detailed phenomenological calculations will be reported elsewhere.
Moreover, one may argue that standard Higgs couples to the 
extra Higgs singlet through $(\phi^{\dag}\phi)(\Phi^{\dag}\Phi)$. It turns 
out, however, that 
the renormalizability is not spoiled by assuming the 
absence of $\phi$-$\Phi$ coupling since fermions do not couple to the 
additional Higgs which do not couple to the usual gauge bosons. Hence, one 
can assume that there exists no $\phi$-$\Phi$ coupling.
\footnote{Nonetheless, 
there is no fundamental reason to prohibit $\phi$-$\Phi$ coupling
from the model.}
Only neutral current interaction due to
$Z'$ exchange contributes at yet unknown very high energy. 
The effective 
neutral coupling due to $Z'$ exchange is given by
\begin{eqnarray}
\cL_{NC}^{eff}&=&-\frac{G'}{\sqrt{2}}J_{Z'\mu}J^\mu_{Z'},\hspace{0.5cm}
G'=\frac{{g'}_1^2\sin^2{\delta}}{2M_{Z'}^2}\ll G=\text{\rm Fermi}\;\text{
\rm constant}.
\label{eqn:34}
\end{eqnarray}
\\
\ind
We have presented
a minimum extension of anomaly-free standard model
with the gauge group $G'=SU(3)\times 
SU(2)\times U(1)\times U(1)$.\cite{4)}
This is made possible if
$\nu_R$ is assumed to be non-singlet under $G'$.
Consequently, $\nu_R$ makes an important
contribution to anomaly cancellation.
A new energy
scale is also introduced,
 which provides mass of 
new extra gauge boson and Majorana mass of $\nu_R$.
\section*{Acknowledgements}
The authors are grateful to
H. Kase for helpful discussions.

\end{document}